 \documentclass[times,10pt,twocolumn]{article}
   \usepackage{latex8}
   \usepackage{times}
\usepackage{url}
\usepackage{amssymb,amsmath,amsthm}
\newtheorem{theorem}{Theorem}[section]

\begin{document}

\title{The Quantum Locker Puzzle \\  \vspace{.2cm}
}

\author{
 David Avis \\ McGill University \\ \url{avis@cs.mcgill.ca}
 \and
 Anne Broadbent\\
IQC, University of Waterloo \\
 \url{albroadb@iqc.ca}
 }

\hyphenation{avenues}
\hyphenation{research}
\hyphenation{interact}
\hyphenation{machine}
\hyphenation{machines}
\hyphenation{giving}
\hyphenation{paper}
\hyphenation{encoun-ter}
\hyphenation{encoun-ters}
\hyphenation{expe-ri-ence}
\hyphenation{analyse}
\hyphenation{analysis}

\clubpenalty 10000
\widowpenalty 10000

\maketitle \thispagestyle{empty}
\begin{abstract}
The \emph{locker puzzle} is a game played by multiple players against a referee.
It has been previously shown that the best strategy that exists cannot succeed with probability greater than~\mbox{$1-\ln2 \approx~0.31$}, no matter how many
players are involved. Our contribution is to show that quantum players can do
much better---they can succeed with \mbox{probability~1}. By making the rules of the
game significantly stricter, we show a scenario where the quantum players still
succeed perfectly, while the classical players win with vanishing probability.
Other variants of the locker puzzle are considered, as well as a cheating
referee.

\end{abstract}

\Section{Introduction}
Grover's quantum algorithm~\cite{Grover} provides a quadratic speedup over the
best possible classical algorithm for the problem of unsorted searching in the
query model.
While Grover's search method has been shown to be optimal~\cite{BBHT96}, our
results reveal that in the context of  multi-player query games, applying
Grover's algorithm yields success probabilities that are much better than the
success probabilities of  classical optimal protocols. Specifically, we show
that in the case of the \emph{locker puzzle}, quantum players succeed with
probability~1 while the known optimal classical success probability is bounded
above by \mbox{$1-\ln2 \approx~0.31$}. In order to amplify this separation, we
prove that a significantly stricter version of the locker puzzle has vanishing
classical success probability, while still admitting a perfect quantum
strategy. We also consider the empty locker and the coloured slips versions of
the locker puzzle, and the possibility of a cheating referee.

\Section{The Locker Puzzle}\label{sec:locker puzzle}

The \emph{locker puzzle}\footnote{The \emph{locker puzzle} also sometimes
refers to another scenario that involves the opening and closing of locker
doors in a hallway, where the question is:  after a specific series of moves,
which locker doors remain open? Our puzzle here is different (and much more
challenging).} is a cooperative game between a team of $n$ players numbered $1,
2, \ldots, n$ and a referee. In the initial phase of the game, the referee
chooses a random permutation $\sigma$ of $1, 2, \ldots, n$, and for each
player $i$
she places number $i$ in locker $\sigma (i)$.
In the following phase, each player is
individually admitted into the locker room. Once in the room, each player is
allowed to open $n/2$ lockers, one at a time, and look at their contents (for
simplicity, we'll take $n$ to be even). After the player leaves the room, all
lockers are closed. The players are initially allowed to discuss strategy, but
once the game starts, they are separated and cannot communicate.  An individual
player~$i$ \emph{wins} if he opens a locker containing number~$i$, while the
team of $n$ players \emph{wins} if all individual players~win.

We would like to know what is the best strategy for the team of $n$ players.
A na\"ive approach is for each player to independently choose $n/2$ lockers to
open. Each players wins independently with probability $1/2$, hence the team
wins with probability $1/{2^n}$. Surprisingly, it is known that the players can do much, much
better! We will review in Section~\ref{section:optimal-classical-locker} an
optimum strategy by which, for
any~$n$, the players can win with probability at least~$0.30685$.


The locker puzzle was originally considered by Peter Bro Miltersen, and was
first published in~\cite{GalMiltersenICALP};
a journal version appears in~\cite{GalMiltersenTCS}.  Sven Skylum is credited
for the pointer-following strategy that we will give in the next section.
A proof of optimality for this strategy is given by
Eugene Curtin and Max Warshauer~\cite{MathIntelligencer}.
Our presentation of the classical puzzle and its solution follows along the
lines of their article.
Many variations have been
proposed~\cite{GoyalSaks} . We will consider the variations of \emph{empty
lockers} in Section~\ref{section:empty-lockers}, \emph{coloured slips} in
Section~\ref{section:coloured-slips} (to be accurate, the locker and the
coloured slips puzzles are variants of the empty locker puzzle),
and a \emph{cheating referee} in Section~\ref{sec:cheating}.

\SubSection{An Optimal Classical Solution}
\label{section:optimal-classical-locker}

We saw that a na\"ive solution allows the players to win with an exponentially
small probability. How can we devise a strategy that does better? The reader
avid to search for a solution on his or her own is encouraged to do so now.

The key is to find a solution where the individual success probabilities are
not independent. Consider the following strategy: when first entering the
locker room, player~$i$ opens locker number~$i$. A number is revealed; this is
used to indicate which locker to open next (i.e.~if number~$j$ is revealed, the
next locker opened is locker~$j$). Each player executes this pointer-following
strategy until $n/2$ lockers are opened.

To analyze the success probability, note that the team will win provided that
the \emph{permutation} that corresponds to the placement of  numbers in lockers
by the referee does not contain a cycle of length longer than $n/2$. The
probability of such a long cycle occurring is:
\begin{equation}
\sum_{k=1}^{n/2} \frac{1}{{n/2}+k}\,.
\end{equation}

It can be shown that as $n \rightarrow  \infty$, $\sum_{k=1}^{n/2}
\frac{1}{{n/2}+k} \rightarrow \ln 2$ and that the sum increases with~$n$. Hence
the probability that the team wins  is decreasing to $1-\ln 2 \approx 0.30685$.

Using a reduction to another game, this strategy can be shown to be optimal
~\cite{MathIntelligencer}.

\Section{A Quantum Solution}
\label{sec:quantum solution}

We now present our first contribution: a quantum solution to the locker puzzle,
which performs better than the classical solution.

As before the referee chooses a random permutation $\sigma$ and she places
numbers in the lockers according to this permutation.
In the quantum solution, we allow the players to open locker doors in
\emph{superposition}, each player working with his own quantum register.  This
is analogous to the quantum query model.
For the quantum case, we need to modify the goal of the game which, for
player~$i$, becomes  to \emph{correctly guess} the locker containing number~$i$
after $n/2$ queries, and \emph{not} to open locker containing number~$i$,
because this would be too easy to do in superposition!  We show that quantum
players can always win at the locker game. In fact, our results are stronger:
we give a stricter version of the locker puzzle for which the optimal classical
solution succeeds with vanishing probability, while a quantum strategy always
succeeds!

\SubSection{Improving the Success Probability}\label{Imp}

The main idea  is to apply Grover's quantum search algorithm to the locker
puzzle. For player~$i$, we consider the action of opening a locker as a query
to the oracle which when input locker number ~$x$,
$1 \le x \le n$, outputs the following:
\begin{equation} \label{eq:oracle}
f_i(x) = \begin{cases}
1 \text{ if }  \sigma(i)=x \\
0 \text{ otherwise\,.}
\end{cases}
\end{equation}
Note that this oracle is weaker than the oracle in the original puzzle which
would output $f_i(x) = \sigma ^ {-1}(x)$. We
discuss this further in
Section \ref{subsection:optimality} and in the conclusion.

Grover's search algorithm \cite{Grover} was thoroughly analyzed in
\cite{BBHT96}, where it was shown that in a black-box search scenario where it
is known that a single solution exists, $\frac{\pi}{4}\sqrt{n}$ queries yield a
failure probability no greater than~$\frac{1}{n}$, where~$n$ is the number of
elements in the search space (here,~$n$ is assumed to be large).
This was further improved in~\cite{BHMT02}, where is was shown that the same
amount of queries is sufficient to find a solution with \emph{certainty}.

Applying this directly to the quantum players of the locker puzzle yields the
following:
\begin{enumerate}
\item  \label{step:GroverQuery}Each player performs  $\frac{\pi}{4}\sqrt{n}$
queries (this is less than the $\frac{n}{2}$ queries in the classical solution).
\item Each player wins independently with certainty, implying that the  team
wins with certainty.
\end{enumerate}

\SubSection{Reducing the Number of Queries}
\label{sec:reducing-number-queries}

We've seen that quantum players of the locker game can succeed with
probability~1. Our solution only requires $\frac{\pi}{4}\sqrt{n}$ oracle
queries per player. Hence, we now consider the asymptotically stricter version of the
locker puzzle, where players are allowed to open at most $\sqrt{n}$ lockers.
The next theorem state that the success probability for classical players goes
quickly to~0.

\begin{theorem}\label{thm:classic}
In the locker puzzle with $\sqrt{n}$ queries, classical players win with
probability at most~$\frac{1}{\lfloor \sqrt{n} \rfloor !}$.
\end{theorem}

\begin{proof}
Let $N= {\lfloor \sqrt{n} \rfloor }$. We upper bound the success probability of the first $N^2$ players,
when each player is allowed to open $N$ lockers. Since $n \ge N^2$, this upper bounds the
success probability of all $n$ players.

Consider a new game
where the first player opens exactly~$N$ lockers and publicly reveals all
of their contents. If the first player's number is not revealed the players lose and the
game is over. Otherwise the $N$ revealed players have successfully located their lockers.
These $N$ lockers and players are now removed from the game.
The first player has success probability at most~$N/N^2$.

In successive rounds, a player is chosen from amongst those
not yet removed from the game.
He continues in the same way by choosing $N$ of the remaining lockers
and revealing their contents.
If he finds his label, again $N$ lockers and players are removed from the game.
The game stops whenever a chosen player does not find his label.
Otherwise it continues for $N$ rounds and terminates with a win for the players.

The success probability of the new game is at most
\begin{equation}
\frac{N}{N^2} \cdot \frac{N}{N^2 -N}\cdot  \frac{N}{N^2-2N} \cdot \ldots \cdot
\frac{N}{N} = \frac{1}{N!}\,. \end{equation}
The original game with no revealing of numbers cannot do better.
\end{proof}

\SubSection{Optimality and Oracle Strength}
\label{subsection:optimality}
\begin{theorem}
In the quantum query model with oracle (\ref{eq:oracle})
the total number of queries required to obtain a success probability of one
for the players is
in $\Omega (n \sqrt{n})$.
\end{theorem}
\begin{proof}
First
consider a variation of the quantum game where the players act sequentially
in the order $1,2, ..., n$ and are allowed to announce their
results to the other players. The number of queries performed by
Player 1 must be in $\Omega (\sqrt{n})$ or he will not succeed with
probability one. This follows from the analysis of Grover's algorithm,
see \cite{BBHT96}.

The only information given by the oracle $f_1$
is the location of the locker containing label $1$.
Suppose player 2 is allowed to receive this information and remove that locker
from consideration. The permutation $\sigma$ induces a random permutation
on the remaining $n-1$ lockers. Player 2's success probability is then one
only if his number of queries is in $\Omega (\sqrt{n-1})$. Continuing,
the $i$-th player must ask a number of queries in $\Omega (\sqrt{n-i})$.
The total number of queries is therefore in~$\Omega (n \sqrt{n})$.

In the modified game we share all information available to all players
that have not already played. So this shows a lower bound of the same
order for the original version of the quantum game where no information is shared.
\end{proof}

Let us now compare the strength of oracle (\ref{eq:oracle})
with the stronger oracle where $f_i(x) = \sigma ^ {-1} (x)$.
In the classical setup, the weaker oracle (\ref{eq:oracle}) merely tells a given player whether
or not his label is in a requested locker. There are an even number $n$ of lockers
and he can ask $t=n/2$ queries.
Again we consider a sequential version of the game
as described above, where each player reveals his results. If he succeeds, he
reveals the locker with his number and that locker is removed.
For the other lockers he queried, the only information he has is that
they did not contain his label. Therefore after his locker is removed,
the other players have no further information.
The success probability of this variation of the locker game
is:
\begin{eqnarray}
\frac{t}{2t} \cdot \frac{t}{2t-1}\cdot  \frac{t}{2t-2} \cdot \ldots \cdot
\frac{t}{t+1} = \frac{t^t t!}{(2t)!} \\ \approx
\frac{t^t \sqrt{2 \pi t }~t^t e^{-t}}{\sqrt{2 \pi 2t }~(2t)^{2t} e^{-2t}}
~=~ \frac{1}{\sqrt{2}} \left( \frac{e}{4} \right) ^ t
\approx \frac{1}{\sqrt{2}} 0.824 ^ n,
\end{eqnarray}
where we have used Stirling's formula twice. This is exponentially small
and provides an upper bound on the success probability of the classical locker game
with the weak oracle (\ref{eq:oracle}).
By comparison, as we saw in Section \ref{section:optimal-classical-locker}
the players can win with constant probability using the stronger oracle.
An open question is whether the quantum algorithm can be improved by
using this stronger oracle.

\Section{Variants of the Locker Puzzle}

The original motivation for the locker puzzle came from the study of time-space
tradeoffs for the substring search problem in the context of \emph{bit probe
complexity}~\cite{GalMiltersenICALP}. There, a version with both \emph{empty
lockers} and \emph{coloured slips} was presented.
We now examine these two variations separately and consider the quantum case.

\SubSection{Empty Lockers}
\label{section:empty-lockers}

Suppose there are a total of $b \geq n$ lockers.
The referee selects an unordered subset $\sigma$ of
$\{ 1, \ldots, b \}$ with cardinality
$n$ and she puts label $i$ into locker $\sigma(i)$ for $i=1, \ldots, n$.
The remaining $b-n$ lockers are empty.
Assume $b$ is even, and we allow the players to open up to $b/2$ lockers.
An optimum winning strategy for
this more general situation is unknown: the pointer algorithm fails
if an empty locker
is opened.
Even for the case $b=2n$, where half of the lockers are empty,
it is still unknown if there is a classical strategy with success
probability bounded away from zero \cite{GoyalSaks}.
However, the quantum strategy given in
Section~\ref{sec:quantum solution} still succeeds with probability one with a number of
queries in~$O(\sqrt{b})$ per player, for a total of $O( n \sqrt {b})$ queries.
It suffices to modify the oracle
(\ref{eq:oracle}) so that $x$ runs over the range $1 \leq x \leq b$,
and query it $\frac{\pi}{4}\sqrt{b}$ times.
If it turns out that for these same parameters,  the
classical success probability vanishes, then the power of the quantum world
would be once more confirmed, as in Section~\ref{sec:reducing-number-queries}.
and  Section~\ref{subsection:optimality}.

\SubSection{Coloured Slips}
\label{section:coloured-slips}

Consider the empty lockers game with $b \geq n$ lockers, again with
$n$ players and  $n$ slips
of paper, each labelled $1, \ldots, n$. This time the referee colours
each slip
either red or blue as she chooses, and places them in a randomly selected
subset of $n$ lockers.
As before,  each player~$i$ may open $b/2$ lockers using
any adaptive strategy, and based on this, must make a guess about the colour of
the slip labelled~$i$. The players win if every player correctly announces the
colour of his slip. With $b=n$, this can be solved with the pointer-following algorithm
and the players have success probability about 0.31.

In the quantum setting, the players can win with probability one at
the colour guessing game also, by changing the oracle (\ref{eq:oracle}).
Let $c(i)$ be the colour of the slip for player $i$.
Define for $1 \leq x \leq b$ and
$1 \leq i \leq n$:
\begin{equation} \label{eq:oracle1}
g_i(x) = \begin{cases}
1 \text{ if }  \sigma(x)=i~\text{ and }~c(i)=red\\
0 \text{ otherwise.}
\end{cases}
\end{equation}

Now we use the protocol described in Section \ref{Imp} with each player
querying this new oracle $\frac{\pi}{4}\sqrt{b}$ times.
If for player $i$ $c(i)=red$, then there is exactly one
$x$ for which $g_i (x) =1$ and Grover's algorithm returns $x = \sigma (i)$
with probability one.
Otherwise, if $c(i)=blue$ then $g_i$ is identically zero and Grover's algorithm
may return any value $x$. The player now makes one further call to
oracle (\ref{eq:oracle1}) with the returned value $x$ and guesses
red if the oracle returns one and blue
otherwise.

\SubSection{Cheating Referee} \label{sec:cheating}

A cheating referee can obviously beat the players in the locker game.
She simply has to omit the label of one of the players.
This could be easily exposed by requiring that all the lockers be
opened and checked at the end of the game.

A more subtle way of cheating is if the referee
can somehow choose the permutation
$\sigma$. In the original locker game,
let $s = n/2 + 2$, and let $i_1 , \ldots , i_s$ be a random
unordered subset of $s$ players. She may set
$\sigma (i_1) = i_s $, $\sigma (i_{j+1}) = i_j , j=1, \ldots , s-1$,
and fill out the rest of $\sigma$ at random from the remaining players.
It is easy to verify that, using the pointer algorithm,
player $i_1$ opens $n/2$ lockers
$i_1 , \ldots , i_{s-2}$ and does not find his label.
He has to guess and loses with probability about $2/n$.
The same thing happens for each of the players $i_j$.
(Incidentally, the reason for not choosing $s=n/2+1$
is that the players not finding their label may guess the locker number they see
in the last locker they open, winning the game with probability one!).
Using variants of this idea the referee may cheat successfully
for some time before the players catch on. If the players have access to shared randomness (which is unknown to the referee), they can circumvent this problem by first applying their own permutation on the lockers before opening any of them. Interestingly, our quantum protocol is impervious to a referee who maliciously chooses the permutation, and does not require shared randomness.

\Section{Conclusion and Discussion} \label{sec:conclusion}

It was previously known that the locker puzzle has an intriguing classical
optimal solution. Now we know that the locker puzzle and its variants also have
interesting quantum solutions which perform significantly better than the classical ones.

We have given a quantum solution in the black-box query complexity model that
\emph{does not use the pointer-following technique that is crucial to the
classical optimal solution}. It would be interesting to see if using the
stronger classical oracle could lead to a quantum solution that works with a
reasonable probability of success using~$o(n \sqrt{n})$
total queries.
With this stronger oracle,
perhaps shared entanglement could help the
players?
It would also be interesting to see if, analogous to the classical case,  our
results have any consequences for time-space tradeoffs for data
structures~\cite{GoyalSaks}.

\section*{Acknowledgements}
 We would like to thank Bruce Reed for introducing us to the classical version
of the locker puzzle and Richard Cleve
for pointing out the perfect quantum
search of~\cite{BHMT02}.  This work was partially supported by an
an NSERC discovery grant and an NSERC postdoctoral fellowship.

\bibliographystyle{latex8}

\end{document}